# High-performance controllable ambipolar infrared phototransistors based on graphene-quantum dot hybrid

Ran Wang[1,2], Yating Zhang[1,2,*], Haiyang Wang[1,2], Xiaoxian Song[1,2], Lufan Jin[1,2], Haitao Dai[3], Sen Wu[1], and Jianquan Yao[1,2]

[1] Institute of Laser & Opto-Electronics, College of Precision Instruments and Opto-electronics Engineering, Tianjin University, Tianjin 300072, China

[2] Key Laboratory of Opto-electronics Information Technology (Tianjin University), Ministry of Education, Tianjin 300072, China

[3] Tianjin Key Laboratory of Low Dimensional Materials Physics and Preparing Technology, School of Science, Tianjin University, Tianjin 300072, China

Corresponding author: yating@tju.edu.cn

**The field effect transistors (FETs) exhibited ultrahigh responsivity ($10^7$ A/W) to infrared light with great improvement of mobility in graphene / PbS quantum dot (QD) hybrid. These reported transistors are either unipolar or depletion mode devices. In this paper, we presented and fabricated conveniently-controlled grapheme / PbS QD hybrid FETs. Through the investigation on electric and optoelectronic properties, the ambipolar FETs (normally OFF) can be switched ON by raising gate voltage ($V_G$) up to 3.7 V and -0.8 V in the first and third quadrants. Near these thresholds ($V_T$) each carrier species shows comparable mobility (~ 300 $cm^2V^{-1}s^{-1}$). Photo-responsivity reach ~ $10^7$ A/W near each threshold and it will linearly increases with ($V_G$-$V_T$). These hybrid FETs become strongly competitive candidates for devices in flexible integrated circuits with low cost, large area, low-energy consumption and high performances.**

Recently, a breakthrough of graphene-quantum dot hybrid field effect transistor (FET) in photo-responsivity has attracted much attention [1, 2]. Taking advantages of the extreme high mobility of graphene and strong light-absorbing properties of PbS quantum dots (QDs), hybrid metal oxide semiconductor field effect transistors (MOSFETs) exhibit excellent performance in photo-responsivity and gain [1, 2]. The reported responsivities are as high as $10^7$ $AW^{-1}$, while the gain up to $10^8$, which is at least seven order of magnitude larger than graphene FETs or QD FETs that had been reported before [3-10]. The sensitivity and spectral selectivity can be tuned by altering the gate and size of QDs, respectively. However, the operational gate voltage is from -40 V to 80 V [1, 2], which is too large for a flexible integrated circuit. Low operational voltages are of importance in meeting the requirements of low-energy consumption and portable optoelectronic products. Moreover, as building blocks, ambipolar FETs show wider applications than monopolar ones, for type of those is also switched by gate as required. However, n-type FETs are not easily achieved as p-type ones based on solution processed colloidal QDs [11]. On the other hand, near-infrared detectors are dominated by III-V semiconductors. Lower temperature and higher bias voltage are also required for wider spectral response or higher responsivity.

Here, we presented and fabricated FETs based on graphene-PbS QD hybrid, which exhibits high performance at zero bias voltage and low gate threshold. By tuning gate, FETs can be switched among ON mode, OFF mode and amplification mode. Moreover, they work both in the first and the third quadrants where electrons and holes exhibit comparable ultra-high mobility on the order of sub 1000 $cm^2V^{-1}s^{-1}$. Under zero bias voltage, the photoelectrical responsivity reaches as high as $10^7$ A/W. The width of spectral response is more than 1200 nm. Therefore, controllable ambipolar FETs with ultra-low threshold and high responsivity are strongly competitive candidates to building blocks in flexible integrated circuits with low cost, large area, and performance comparable to traditional semiconductor FETs.

As mentioned above, the hybrid FETs based on graphene and PbS QDs have exhibited excellent performance in ultra-high photoresponsivity. Due to the extraordinary high mobility of graphene layer (up to 60,000 $cm^2$/Vs), the hybrid FETs provide a channel for carrier transportation; QDs layers also offer sufficient photo-induced carriers to graphene. The performance of hybrid FET strongly depends on the cooperation of graphene and QDs, especially energy matching of two components. Therefore, to achieve low operational voltage, attempts need to be paid on making Fermi level of PbS QDs close to 4.6 eV in accordance with Dirac point of graphene. As a result, QDs of 5 nm in diameter were obtained by colloidal chemical method; see *Supplementary Information* (SI).

PbS is a semiconductor with narrow bandgap of 0.41 eV (at 300 K) and with electron affinity of ~ 4.7 eV when it is a bulk material [12, 13]. As the spacing scale of crystal shrinking down to several nanometers, bandgap increases due to quantum confinement effect [12]. According to absorption peak (at 1200 nm), the bandgap could reach 1.03 eV, and the average diameter could be estimated to be 5 nm with consistence of statistical results on transmission electron microscope (TEM) [14]. Due to nearly equal effective masses of electrons and holes in PbS, the conduction band and the valence band levels of QDs are estimated to be 4.2 eV and 5.2 eV, respectively, as shown in Fig. 1 (a). Notably, whether PbS QDs is p-type or n-type semiconductor strongly depends on surface ligands [15-17]. Experimentally, we used ethanedithiol (EDT) to treat PbS QDs. Therefore, QD film can be regard as an n-type semiconductor layer [6, 17].

The structure of FETs based on graphene and PbS QDs has been plotted in Fig. 1 (b). A single layer of graphene sheet was transferred onto an $n^+$ Si / $SiO_2$ substrate. The thickness of $SiO_2$ is 300 nm. The source and drain electrodes were deposited on top of graphene $A_l$ over a shadow mask by thermal evaporation. Then, three layers of PbS QDs were deposited by layer-by-layer (LBL) approach from toluene solution via spincasting and the details of the approach are described in SI.

Electronic and photoelectronic measurements are performed according to the circuit diagram as shown in Fig. 1. Bias $V_{SD}$ is applied over source (with ground) and drain electrodes by Keithley 2400, channel current flowing into drain denoted by $I_D$ is measured by Keithley 2400. Voltage $V_G$ is applied on gate by HP6030A with reference to ground.

If QDs come in direct contact with graphene layer, build-in electric field will set up at the interface. However, the as-prepared QDs are capped with surface ligand, such as oleylamine (OLA) experimentally. Hence, carriers who transfer from QDs to graphene have to cross the ligand barrier. Recent studies show that the width of barrier plays more important role than barrier height [6]. The decrease of carbon chain length of ligand can narrow QDs, which increases effective transfer rate of carriers. Experimentally, OLA was replaced by EDT. Even if EDT is much shorter than OLA, carriers are not free to cross the interface. As an effective method, a series of short voltage pulses have been applied over drain to source and gate (see SI). Repeating that process for several times, a stable charge channel will set up from source to drain. As the same time, charge transformation over the interface between graphene layer and PbS QD layer shows that p- or n- type of hybrid mainly depends on the doping type of graphene. So that type of FET could be switched by adjusting carrier concentration in graphene via gate.

Subsequently, electrical properties are characterized by output and transfer characteristics. Fig.2 shows gate dependence of output characteristics in the first (a) and the third (b) quadrants. For protection, current $I_D$ is limit within ±1 mA. It is a normally OFF switch, n- or p- channel depends on gate voltage. When negative (positive) gate voltage is applied, electrons (holes) in graphene cross the interface barrier and flow over into the QDs, remaining holes (electrons); hence p- (n-) channel is built in graphene layer. With the increase of negative (positive) terminal bias, drain current $I_D$ transfers from linear region to saturation region.

For low value of gate $V_G$ and bias voltage $V_{SD}$, such as zero volts, a small number of carriers transfers into QDs while an equivalent number of carriers remains in graphene and pin on all channels. With the bias voltage increasing from zero, current $I_D$ increases linearly due to higher transfer velocity from higher electric field. In this region, the mobility keeps as a constant. As the bias voltage reaches to a value $V_{DS}$, the current remains at a constant value in the saturation region due to the pinning-off effect, in which the remained carriers occupied a part of channel. Both holes and electrons exhibit such typical FET characteristics, so the same mechanism plays a role. With the increase of $V_G$, the increasing number of carriers reduces higher value of $V_{SD}$ and saturated current.

Fig.2 shows bias dependence of transfer characteristics in the first (c) and the third (d) quadrants. When $V_{SD} \ll V_G$, transfer characteristics are in linear region with the constant mobility. The expression of mobility is given by [18, 19]

$$\mu = \frac{L}{WC_{ox}(V_G - V_T)} \cdot \frac{I_D}{V_{SD}} \quad (1)$$

Where W and L is width and length of the channel, respectively. $C_{ox}$ is capacitance of gate dielectric per unit area, and second term of the right is the inverse of the sloop of low $V_{SD}$ characteristics. Then mobility of electrons and holes are calculated out to be 295 and 333 $cm^2V^{-1}s^{-1}$, see SI.

With the increase of $V_{SD}$, an extreme value of $I_D$ emerges at a point in the first and the third quadrants, which are the critical point of species of carriers, termed to $V_{GC}$.

Then in the same condition we measured output and transfer characteristics under laser irradiation of 7.5 mW/cm$^2$ with wavelength of 808 nm as shown in Fig. 3. The profiles of characteristics show similar to those under dark. Data with and without laser irradiation are plotted in Fig. 4 (a) under the same bias $V_{SD} = \pm 5V$ for the comparison with each other. It is clear to see that when exposed to laser, the transfer characteristics hop along the direction of gate increase.

Excitons (electron-hole pairs) generate due to the absorption of the incident light by QDs. Additional electrons (or holes) separated by gate go across the interface increase channel current $I_D$. Compared with the case without light excitation, the role of photo generated carriers can be regarded as an additional induced gate voltage which could be called "light-gate effect". Due to that effect, horizontal shifts $\Delta V_G$ can be used for FET to calibrate optical power received (P) response, as [1]

$$\Delta V_G = \alpha P^\beta \quad (2)$$

Where α and β are constants. P can be directly determined according to equation (2), extraction of $\Delta V_G$ is still not easy. The functional relationship between $\Delta V_G$ and $V_{SD}$ could be obtained by equation $\Delta V_G = V_{GC}^{light} - V_{GC}^{dark}$. It is worth noting that in Fig. 4 (b) $\Delta V_G$ remain undetermined in low bias region, due to absence of $V_{GC}$ in the first and third quadrants. Two distinct sharp decreases of $\Delta V_G$ are shown in those of two quadrants when $V_{SD} = 4$ V and $V_{SD} = -1$ V.

To get more information of response in small $V_{SD}$ region, responsivity R versus $V_G$ is also collected and calculated by [20]

$$R = \frac{I_{ill} - I_{Dark}}{P} = \frac{\Delta I_D}{P} \quad (3)$$

Where $I_{ill}$ and $I_{Dark}$ are channel current under light illumination and in dark, respectively. For example, we select two pairs of curves (in small value $V_{SD} = -1.5$ V, 0.06 V; and large value $V_{SD} = \pm 5$ V) to take compare. The results are plotted in Fig. 4 (c). Obviously, when $V_{SD} = \pm 5$ V, two distinct sharp decreases of R present at I and III quadrants intersecting with horizontal axis at $V_G = -1$V, -0.2 V, 2.7 V and 3.9 V, where light responsivity equals to zero. Two extreme points emerge at -0.5 V and 3.4 V, respectively. That curve trends consist with the functional relationship between $\Delta V_G$ and $V_{SD}$, indicating R is not a monotone function of $V_G$ in large value of $V_{SD}$ region. Thus, operating mode of FET must be carefully selected for optical power detection.

The physical mechanism for the phenomenon above lies in the principles of FET. Increments of channel current $\Delta I_D$, resulting from the light illumination, could be expressed as a function of the gate shift $\Delta V_G$ derivated from equation (1):

$$\Delta I_D = \frac{W}{L} C_{ox} \mu V_{SD} \Delta V_G \quad (4)$$

For one device, $WC_{ox}\mu/L$ is constant, and $\Delta I_D$ presents directly proportion to $V_{SD}\Delta V_G$.

Substituting equation (4) into (3), and we get the expression

$$R = \frac{W}{L}\frac{C_{ox}\mu}{P}V_{SD}\Delta V_G \quad (5)$$

When P is constant, R presents directly proportion to $V_{SD}\Delta V_G$. Hence, Fig. 3 (b) and (c) shows consistent tendency, where an extreme point domains. Here, $V_{SD}$ can be regarded as an amplification factor to $\Delta V_G$, and photo-induced channel current can be enhanced as the price of additional complexity. Moreover, under that mode FET cannot be OFF. When small value of $V_{SD}$ is applied, the switch character is cleanly exhibited. Responsivity can be turn OFF, turn ON, and amplified by $V_G$, threshold gates are -0.5 V and 3.7 V in the third and first quadrant, respectively. Synthesizing each mode of operation, the best choice is low $V_{SD}$ then operating modes are completely controlled by gate.

For a better understanding on relation of sense mechanism and operating mode of FET, we characterized responsivity under light irradiation of difference optical power, when FET operates at zero bias voltage and -1 V gate, as shown in Fig. 5. Responsivity as a function of received optical power P can be obtained by substituting equation (2) into (5) [8, 20],

$$R = \frac{\alpha W C_{ox}\mu V_{DS}}{LA^\beta}P^{\beta-1} \quad (6)$$

In a double-logarithmic axis plot, a good fitting results is achieved, using equation (6). When operates near the threshold, lgR keeps a good linear relation with (β-1)lgP. Moreover, it is sensitive to low optical power in orders of nW. To compare with other work, we also plotted results of R vs P reported by Sun et al[20]. Under same received optical power, R of this work improve almost one order of magnitude than they reported, even operated near threshold. In amplifier region (VG over threshold a lot) responsivity will be further improved.

Next, the transient behavior of the FET was characterized by an optical switch characterization. Variable of channel current $\Delta I_D$ dependence on light were plotted in Fig. 5 (b). FET works over the threshold, and remarkable changes on channel current reaches 22 μA, under illumination 7.5 mW/cm$^2$. The relaxation can be fitted by

empirical expression

$$\Delta I_{SD} = \begin{cases} \Delta I_1(1-\exp(-t/\tau_1)) + \Delta I_2(1-\exp(-t/\tau_2)) & \text{On} \\ \Delta I_3 \exp(-t/\tau_3) + \Delta I_4 \exp(-t/\tau_4) & \text{Off} \end{cases} \quad (7)$$

Where $\tau_1$ and $\tau_2$ are characteristic parameters in response time for laser on, $\tau_3$ and $\tau_4$ are parameters for laser off. Using equation (7), a good fitting consistence are obtained, $\tau_1$ = 0.7 s and $\tau_2$ = 4.5 s. Under condition of $V_{SD}$ = 0 V, $V_G$ = -1 V, $\tau_1$ represents hole transfer time from QDs to graphene while $\tau_2$ represents electron transfer time from source to drain by QDs array. There are two pathways for photo generated carriers, electrons driven to graphene, meanwhile holes forced away from interface. Then they simultaneous drift to drain on respective way, both of which contribute to channel current. Steady state is reached when no more current changes. Then light switched off, concentration of carriers decreases immediately, leading to current drop. $\Delta I_D$ can also fitting by exponential equation(7), $\tau_3$ = 1.8 s and $\tau_2$ = 7.9 s. They are as twice as $\tau_1$ and $\tau_2$, indicating less efficient of light off response than light on. Here, we suppose the rise time $\tau_1$ and decay time $\tau_3$ are holes and electrons transfer time from PbS QDs to graphene, respectively. While rise time $\tau_2$ and decay time $\tau_4$ are electrons and holes transport in PbS QDs.

Fig. 5 (b) shows a light on-off response of channel current. The response of the phototransistor is similar even after device was switched for more than an one hundred times (see SI). The response of the phototransistor is much faster than the photoconductors based on graphene oxide reported before. Since the response time is related to transfer rate of electrons and holes from PbS QDs to graphene, the nature of ligand, capping QDs are essential to performance of the FET.

At last, optical spectrum response of the FET at $V_{SD}$ = 0 V, $V_G$ = -1 V, is measured and compared with optical absorbance of PbS toluene solution, as shown in Fig. 6. In solution, the absorption peaks of PbS QDs locate at mainly 1170 nm, and 1380 nm (side peak). The absorption edge is attributed to 1400 nm, which corresponds to response spectral edge. For photons with higher energy than that, the FET shows a wide response spectral range. The reason lies in that PbS QDs are multiple exciton generators, which means when PbS QD absorbs one photons with higher energy it will generates two or more excitons [21, 22]. Therefore, the FET shows a wide spectral range of response. During the test, no more amplification is used and the change of output current is in orders of milliamp or sub-milliamp.

In summary, by a facile solution process method, controllable ambipolar FET infrared detectors were fabricated. Bipolarity was realized by preparation of PbS QDs with energy structure matching with that of graphene. And good electric and photoelectric performances of hybrid FET are exhibited. Light responsivity can be switched by gate among OFF, ON, and amplifier modes. Moreover, type of FET is in selection by positive or negative of gate. In first and the third quadrants, ultra-low threshold voltages are shown at 3.7 V and -0.8 V, respectively. Responsivity decreases with irradiation increasing. At low irradiation, it reaches to $2*10^5$ A/W when $V_{SD}$ = 0 V, $V_G$ = -1 V. Wide spectral response range is also observed, due to avalanche effect of PbS QDs.

## Acknowledgement

We thank Professor Chunlei Guo from Rochester University for discussion. This work is supported by the National Natural Science Foundation of China (Grant No. 61271066) and the Foundation of Independent innovation of Tianjin University (Grant No. 60302070)

## Reference

1. Sun, Z., et al., *Infrared Photodetectors Based on CVD-Grown Graphene and PbS Quantum Dots with Ultrahigh Responsivity.* Advanced Materials, 2012. **24**(43): p. 5878-5883.
2. Konstantatos, G., et al., *Hybrid graphene-quantum dot phototransistors with ultrahigh gain.* Nat Nano, 2012. **7**(6): p. 363-368.
3. Church, C.P., et al., *Quantum dot Ge/TiO2 heterojunction photoconductor fabrication and performance.* Applied Physics Letters, 2013. **103**(22).
4. Kang, M.S., et al., *High Carrier Densities Achieved at Low Voltages in Ambipolar PbSe*

*Nanocrystal Thin-Film Transistors.* Nano Letters, 2009. **9**(11): p. 3848-3852.

5. Kang, M.S., et al., *Size- and Temperature-Dependent Charge Transport in PbSe Nanocrystal Thin Films.* Nano Letters, 2011. **11**(9): p. 3887-3892.

6. Liu, Y., et al., *Robust, Functional Nanocrystal Solids by Infilling with Atomic Layer Deposition.* Nano Letters, 2011. **11**(12): p. 5349-5355.

7. Liu, Y., et al., *Dependence of Carrier Mobility on Nanocrystal Size and Ligand Length in PbSe Nanocrystal Solids.* Nano Letters, 2010. **10**(5): p. 1960-1969.

8. Konstantatos, G., et al., *Ultrasensitive solution-cast quantum dot photodetectors.* Nature, 2006. **442**(7099): p. 180-183.

9. Reddy, D., et al., *Graphene field-effect transistors.* Journal of Physics D-Applied Physics, 2011. **44**(31).

10. Yigen, S., et al., *Electronic thermal conductivity measurements in intrinsic graphene.* Physical Review B, 2013. **87**(24).

11. Talapin, D.V., et al., *Prospects of Colloidal Nanocrystals for Electronic and Optoelectronic Applications.* Chemical Reviews, 2010. **110**(1): p. 389-458.

12. Kang, I. and F.W. Wise, *Electronic structure and optical properties of PbS and PbSe quantum dots.* Journal of the Optical Society of America B, 1997. **14**(7): p. 1632-1646.

13. Choi, H., et al., *Steric-Hindrance-Driven Shape Transition in PbS Quantum Dots: Understanding Size-Dependent Stability.* Journal of the American Chemical Society, 2013. **135**(14): p. 5278-5281.

14. Cademartiri, L., et al., *Size-dependent extinction coefficients of PbS quantum dots.* Journal of the American Chemical Society, 2006. **128**(31): p. 10337-10346.

15. Talapin, D.V. and C.B. Murray, *PbSe nanocrystal solids for n- and p-channel thin film field-effect transistors.* Science, 2005. **310**(5745): p. 86-89.

16. Koh, W.-k., et al., *Heavily doped n-type PbSe and PbS nanocrystals using ground-state charge transfer from cobaltocene.* Sci. Rep., 2013. **3**.

17. Zarghami, M.H., et al., *p-Type PbSe and PbS Quantum Dot Solids Prepared with Short-Chain Acids and Diacids.* Acs Nano, 2010. **4**(4): p. 2475-2485.

18. Sze, S.M. and K.K. Ng, eds. *Physics of Semiconductor Devices*. third ed. Vol. Chapter 6. 2007, A JOHN WILEY & SONS, JNC., PUBLICATION: Canada. 293.

19. Neamen, D.A., *Semiconductor Physics and Devices: Basic Principles*. Third ed, ed. D.A. Neamen. Vol. Chapter 11 and Chapter 12. 2003: Elizabeth A. Jones.

20. Sun, Z., J. Li, and F. Yan, *Highly sensitive organic near-infrared phototransistors based on poly (3-hexylthiophene) and PbS quantum dots.* Journal of Materials Chemistry, 2012. **22**(40): p. 21673-21678.

21. Yang, Y. and T. Lian, *Multiple exciton dissociation and hot electron extraction by ultrafast interfacial electron transfer from PbS QDs.* Coordination Chemistry Reviews, 2014. **263**: p. 229-238.

22. Yang, Y., W. Rodriguez-Cordoba, and T. Lian, *Multiple Exciton Generation and Dissociation in PbS Quantum Dot-Electron Acceptor Complexes.* Nano Letters, 2012. **12**(8): p. 4235-4241.

Fig. 1 (a) A schematic diagram for charge generation at interfaces of PbS QDs and graphene heterojunction under light illumination, and positive gate. Incident photons create electron-hole pairs in PbS QDs. Electrons (or holes) under positive (or negative) bias are then transferred to the graphene channel and drift towards the drain whereas holes (or electrons) remain in the PbS QDs. (b) A schematic diagram of the graphene-QD hybrid phototransistor: a graphene flake is deposited onto an $n^+$ Si / $SiO_2$ structure overcoated by PbS QD array. Bias ($V_{SD}$) is controlled by Keithley 2400 which also detects the channel current $I_D$, gate voltage is applied by HP 6030A referring to the ground.

Fig. 2 Output characteristics of hybrid FET in the third (a) and first (b) quadrants at dark. Transfer characteristics of hybrid FET in the third (c) and first (d) quadrants at dark at $V_{SD}$ interval of 0.1 V.

Fig. 3 Output characteristics of hybrid FET in the third (a) and first (b) quadrants under laser irradiation of 7.5 mW/cm$^2$ with wavelength of 808 nm. Transfer characteristics of hybrid FET in the third (c) and first (d) quadrants at dark at $V_{SD}$ interval of 0.1 V under laser irradiation of 7.5 mW/cm$^2$ with wavelength of 808 nm.

Fig. 4 (a) comparison output characteristics of hybrid FET with and without illumination in the third and first quadrants. (b) light response shift $\Delta V_G$ versus $V_{SD}$ under irradiation of 7.5 mW/cm$^2$ in the third and first quadrants. (c) responsivity versus $V_G$ under irradiation of 7.5 mW/cm$^2$ at difference of $V_{SD}$ of -1.5 V, 0.06 V, -5 V, and 5 V.

Fig. 5 (a) Responsivity of hybrid FET versus received light power, under operating parameters of $V_{SD}$ = 0 V, and $V_G$ = -1 V (red solid points), fitting line using equation (6) with $\beta$ = 0.0043 (red dashed line); comparison responsivity reported by Sun *et al.*

[1] (black stars). (b) current response of hybrid FET to ON/OFF illumination of 7.5 mW/ cm$^2$ at wavelength of 808 nm, when operates at $V_{SD}$ = 0 V, $V_G$ = -1 V. Fitting line using equation (7), with parameters of $\tau_1$ = 0.7 s, $\tau_2$ = 4.5 s, $\tau_3$ = 1.8 s, $\tau_4$ = 7.9 s. (c) response of channel current to a repeatable laser light on and off.

Fig. 6 Responsive current spectrum of hybrid FET under $V_{SD}$ = 0 V, $V_G$ = -1 V (left), incident monochromatic light is from tungsten lamp (24 V, 150 mW). Absorbance spectrum of PbS QDs in toluene (right).

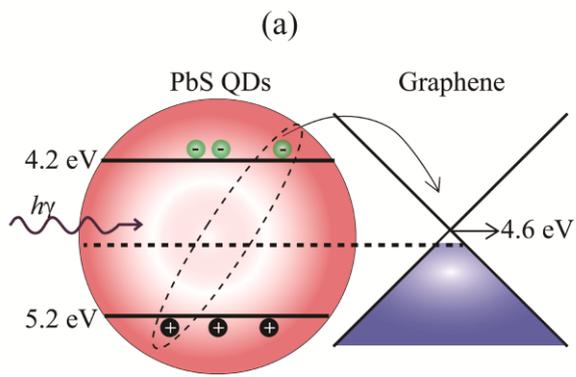 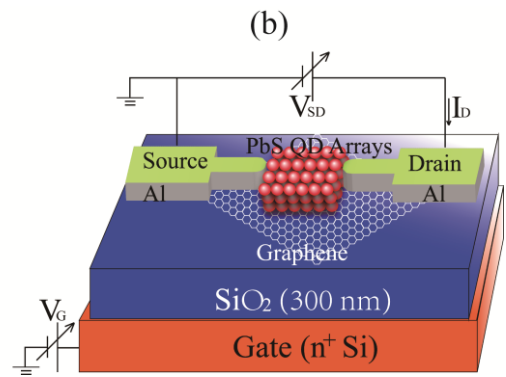

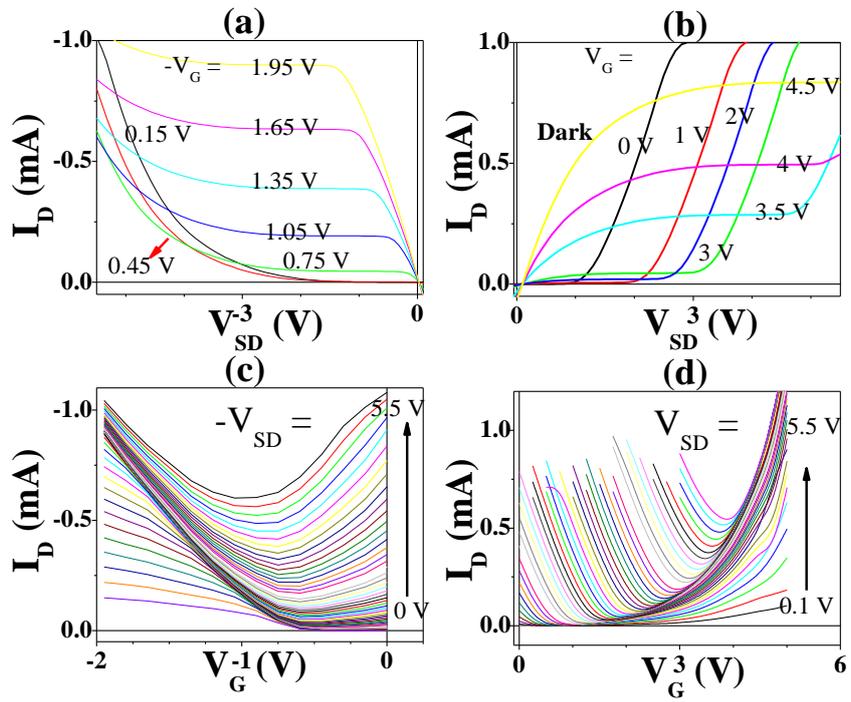

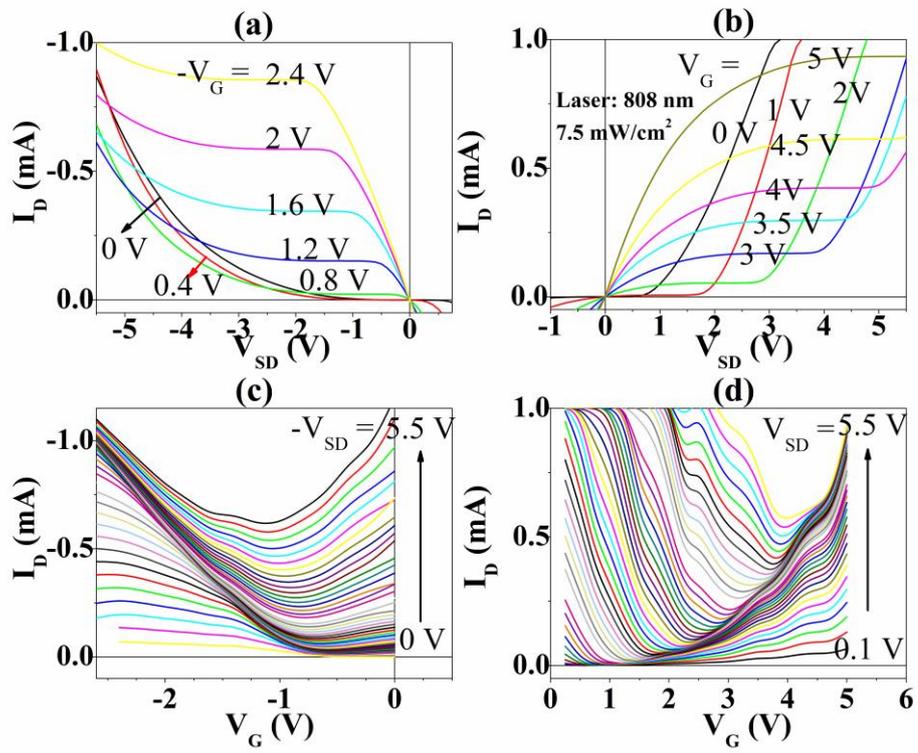

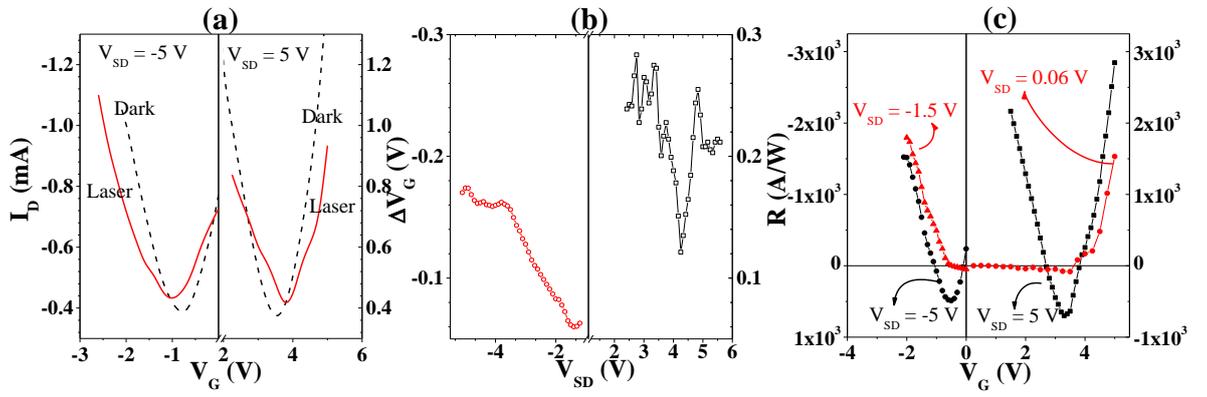

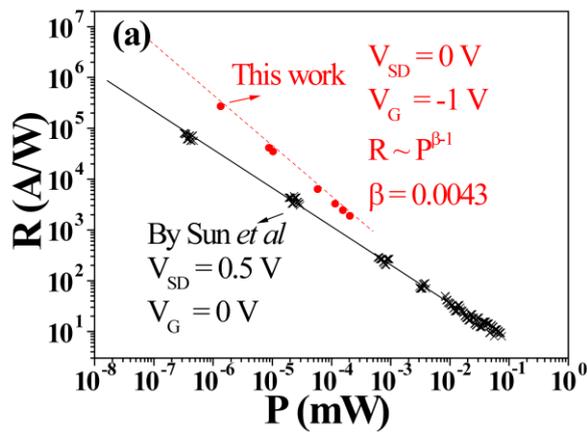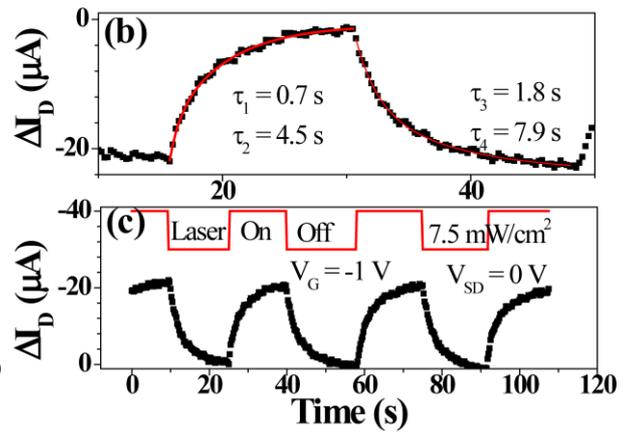

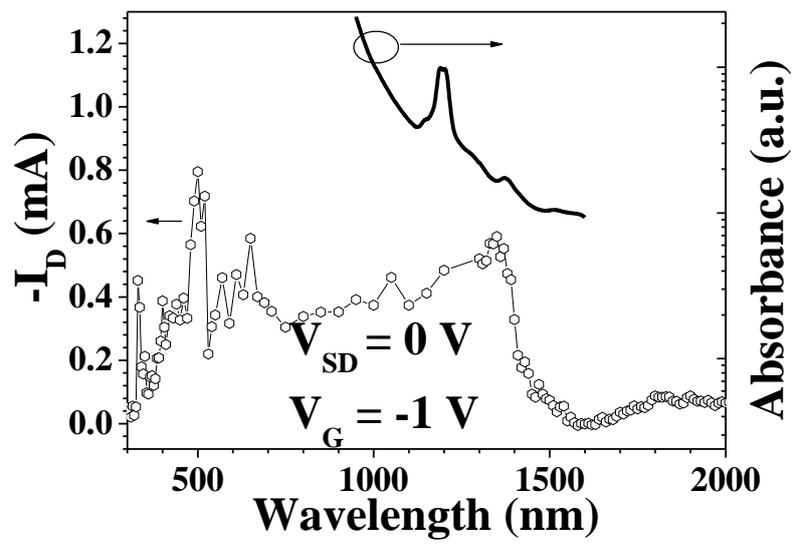